\documentclass[prl,aps,10pt,twocolumn]{revtex4-1}
\usepackage{epsfig,graphicx,amsmath}
\usepackage{amsfonts}
\newcommand{\be}{\begin{equation}}
\newcommand{\ee}{\end{equation}}
\newcommand{\bea}{\begin{eqnarray}}
\newcommand{\eea}{\end{eqnarray}}

\newcommand{\ba}{\bea \begin{array}}
\newcommand{\ea}{\end{array} \eea}
\renewcommand{\(}{\left(}
\renewcommand{\)}{\right)}

\newcommand{\bc}{\begin{center}}
\newcommand{\ec}{\end{center}}
\newcommand{\p}{\partial}

\newcommand{\ds}{\displaystyle}
\newcommand{\beq}{\begin{eqnarray}}
\newcommand{\eeq}{\end{eqnarray}}
\newcommand{\beqq}{\begin{eqnarray*}}
\newcommand{\eeqq}{\end{eqnarray*}}

\newcommand{\eps}{\varepsilon}

\font\bb=msbm10 at 12pt 
 
\def\rR{\hbox{\bb R}}

\newcommand{\vect}[1]{\boldsymbol{#1}}

\begin{document}

%%%%%%%%%%%%%%%%%%%%%%%%%%%%%%%%%%%%%%%%%%%%%%%%%%%%%%%%%%%%%%%%
\title{Computation of the Mean First-Encounter Time Between the Ends of a Polymer Chain}
%%%%%%%%%%%%%%%%%%%%%%%%%%%%%%%%%%%%%%%%%%%%%%%%%%%%%%%%%%%%%%%%
\author{A. Amitai, I. Kupka and D. Holcman}
\affiliation{Group of Computational Biology and Applied Mathematics, Institute of Biology, Ecole Normale Sup\'erieure, 46 rue
d'Ulm 75005 Paris, France.}

\begin{abstract}
Using a novel theoretical approach, we study the mean first encounter time (MFET) between the two ends of a polymer. Previous approaches used various simplifications that reduced the complexity of the problem, leading, however to incompatible results.  We construct here for the first time a general theory that allows us to compute the MFET.  The method is based on estimating the mean time for a Brownian particle to reach a narrow domain in the polymer configuration space. In dimension two and three, we find that the MFET depends mainly on the first eigenvalue of the associated Fokker-Planck operator and provide precise estimates that are confirmed by Brownian simulations. Interestingly, although many time scales are involved in the encounter process, its distribution can be well approximated by a single exponential, which has several consequences for modeling chromosome dynamics in the nucleus. Another application of our result is computing the mean time for a DNA molecule to form a closed loop (when its two ends meet for the first time).
\end{abstract}

\maketitle

\begin{figure}
    \centering
        \includegraphics[width=0.5\textwidth]{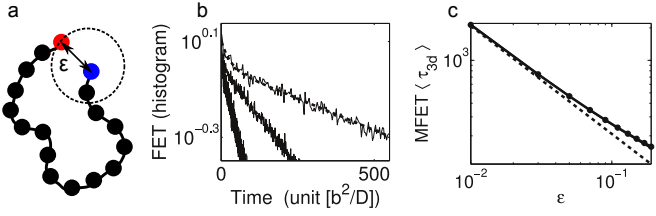}
    \caption{{ First end-to-end encounter time for a Rouse polymer} (a) Scheme of a polymer loop: the two ends are located at a distance of $\varepsilon$ from one another. (b) Histogram of the first encounter times (FETs) obtained from Brownian simulations in three dimensions (full line) and fitting with two exponentials (dashed line) for $N=16,32,64$ (left to right) and $\varepsilon=0.1b$. One exponential is enough for small $N$ (see next figure), while the full dynamics is well captured by at least two for larger $N$. (c) MFET as a function of the radius $\varepsilon$ in three dimensions. Comparison of the Brownian simulations (full line) with the reciprocal of the {first term in the expansion of the} first eigenvalue [Eq. \eqref{first_eigen_eps}] (dashed line) and the full ansatz [Eq. \eqref{MFLT3d}] (circles).}
    \label{fig:merge3d}
\end{figure}

\begin{figure}
    \centering
        \includegraphics[width=0.5\textwidth]{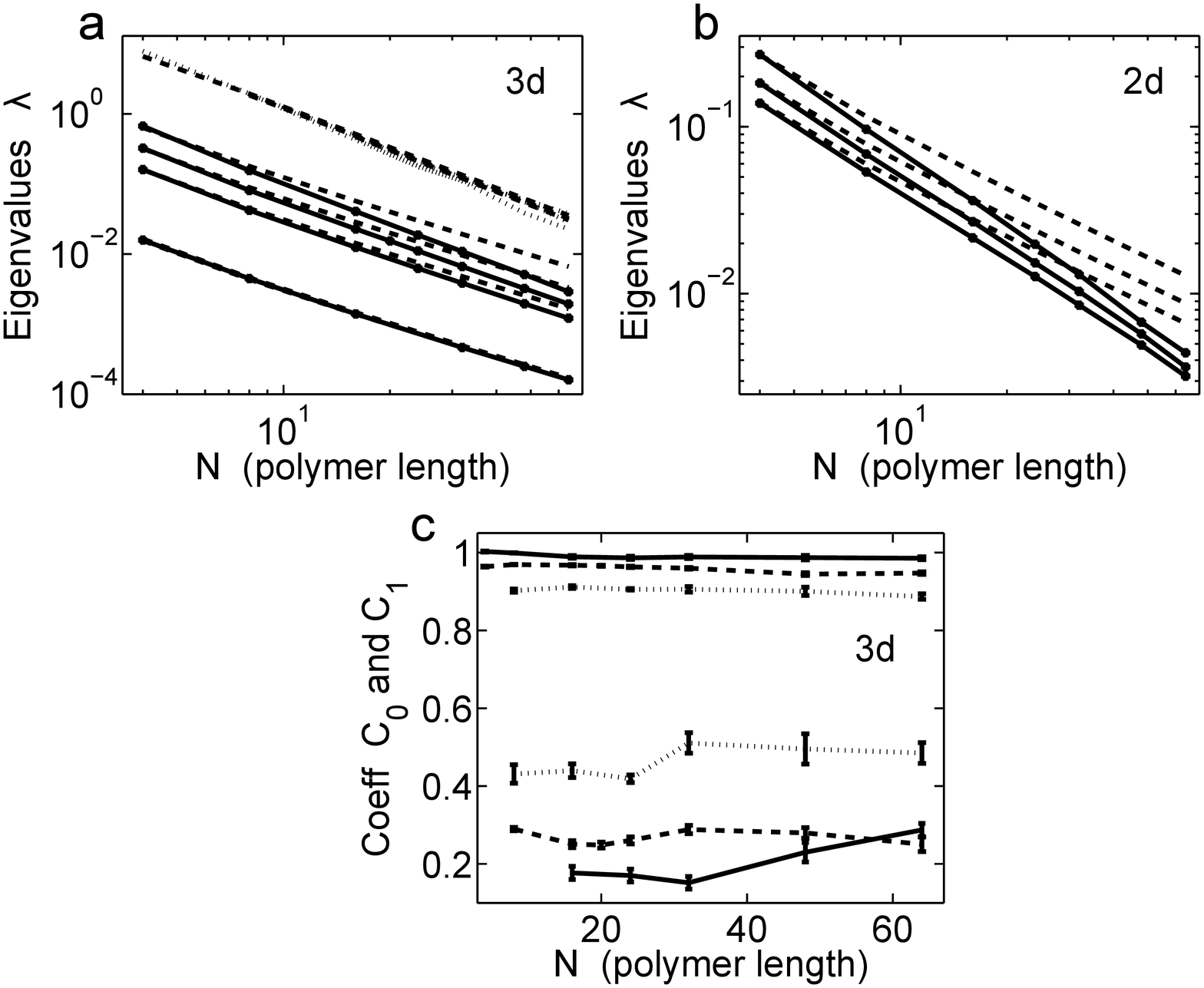}
    \caption{ { The first two eigenvalues of the FET probability [Eq.\ref{FFP2}]}: (a) Using Brownian simulations (full line) in three dimensions and theoretical value, we extract the zero eigenvalue $\lambda^\epsilon_0$ [Eq. \eqref{first_eigen_eps}] (dashed line) for
$\varepsilon=0.01,0.1,0.2,0.4$ (bottom up). The first {nonzero} eigenvalue (cross points) and the theoretical value (dashed line) are computed from eq.\eqref{second_eigen_eps}.
(b) Two dimensional version of (a) for the zero eigenvalue only for $\varepsilon=10^{-4},10^{-3},10^{-2}$ (bottom up). (c) The coefficients $C_0$ and $C_1$
are obtained from Brownian simulations in three dimensions for $\varepsilon= 0.1$ (full
line), $\varepsilon= 0.2$ (dashed line), $\varepsilon= 0.4$ (cross
points) for different polymer lengths. The upper curves corresponds to $C_0$ while the lower ones to $C_1$.}
    \label{fig:fitting_C}
\end{figure}

\begin{figure}
    \centering
        \includegraphics[width=0.5\textwidth]{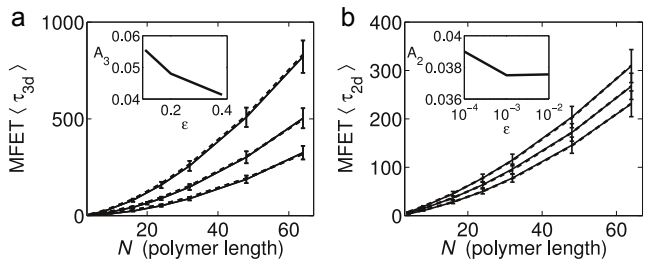}
    \caption{{ Mean first encounter time for different polymer lengths and different values of $\varepsilon$}. (a)  MFET (three dimensions) estimated from Brownian simulations (full line)
and compared to the theoretical MFET [Eq. \eqref{MFLT3d}] (dashed lines). The parameter $A_3$ is obtained by fitting ($\varepsilon=0.1,0.2,0.4$). (b) The MFET (two
dimensions) estimated from Brownian simulation (full lines) and compared to the theoretical MFET [Eq. \eqref{MFLT2d}] (dashed lines). The parameter $A_2$ is obtained by fitting ( $\varepsilon=10^{-4},10^{-3},10^{-2}$).}
    \label{fig:fitting_MFLT}
\end{figure}

The mean time for the two ends of a polymer to meet [Fig. \ref{fig:merge3d}(a)], starting from an open configuration, is a classical and important problem in polymer dynamics that has several implications in DNA looping and in cellular biology where a gene can be activated when a transcription factor bound far away from the promoter site is brought near the active site \cite{Dillon1997,Ptashne1986,Dunn1984}.
Despite much effort both theoretically and numerically \cite{Chen2005, Doi1975_1,Wilemski2,szabo1980,Pastor}, the time scales involved in the formation of a polymer loop by bringing the two ends together remain unclear.\\
The mean first encounter time (MFET) is defined as the first arrival time for the end monomer into a ball of
radius $\varepsilon$, centered at the other polymer end
[Fig. \ref{fig:merge3d}a)]. Interestingly, the MFET does not depend
only on the radius $\varepsilon$ but also on the polymer length $N$
(measured in the number of monomers).
Indeed, an interesting feature of polymer dynamics is the long memory, where the arrival time depends strongly on the initial condition. This property originates from the internal motion of the polymer (characterized by the Rouse modes \cite{Doi:Book}). Specifically, the slowest
relaxation time which is proportional to $N^2$ \cite{Sokolov2003}. The
MFET depends also on the initial end-to-end distribution \cite{szabo1980,Wilemski2,Friedman2006}, a result that was first obtained from a one-dimensional diffusion reduction approach for the end-to-end distance variable.
Thus the MFET depends on the radius $\varepsilon$ and on the slowest relaxation time. Recently, these two time scales were clearly numerically observed \cite{Chen2005,Everaers2012} in a study showing two regimes, depending whether the ratio $\sqrt{N}\varepsilon/b$ is of order 1 or $\gg  1$, where $b$ is the standard deviation of the bond length. In the first regime, the MFET depends on
$\varepsilon$ and scales as $N^{3/2}$, while in the second, it is dominated by $N^2$ {and is independent of $\varepsilon$}.  In summary, the MFET shows mixed scaling laws \cite{Toan2008} with $N$. In addition, in more realistic polymer models such as wormlike-chain model \cite{Hyeon2006} and with hydrodynamical forces, self-avoidance and Coulomb interactions \cite{Podtelezhnikov1997,Stampe2001}, it was shown numerically and using some analytical considerations for the end-to-end distance that the MFET depends on several parameters such as the polymer length and the bending elasticity. However, it is still unclear how to extract the precise dependency with $N$ (scaling law).\\
No systematic approach from first principle was used to derive an expression for the MFET.  As it is a rare event, an analytical formula will facilitate to explore a large fraction of the parameter space, difficult to access numerically or experimentally. Using the  end-to-end distance as a drastic approximation of the dynamics, 
it has been \cite{Pastor} possible to formulate the MFET in terms of a mean first passage time
equation \cite{Schuss:Book}. However, as already noticed "problems
of this type may appear simple but are in fact very difficult
"\cite{Pastor}. Here we undertake this challenge by formulating the MFET as a boundary value problem in the high dimensional polymer configuration space.\\
Our main results consist of formulas for MFET $ \langle \tau_\epsilon \rangle $ in dimension two and three, that account for the two regimes mentioned above and map out the crossover between the two scaling
terms. In addition, we found that the MFET is well approximated by the expansion of the first eigenvalue for the
Fokker-Planck operator associated to the Rouse polymer dynamics. We obtain in dimension two and three, respectively, for small $\varepsilon$ and $N$ such as $\sqrt{N}\varepsilon/b \leq 1$ [see Eq. \eqref{MFET_eign2}],
\begin{subequations}\label{MFLT2d3d}
\begin{align}
 \langle \tau_\varepsilon \rangle_{2d} &=& \frac{N}{2D\kappa}\log\left(\frac{\sqrt{2}b}{\varepsilon}\right) + A_2 \frac{b^2}{D} N^2+\mathcal O(1), \label{MFLT2d}\\
 \langle \tau_\varepsilon \rangle_{3d} &=& \left( \frac{N\pi}{\kappa}\right)^{3/2}\frac{\sqrt{2}}{D 4\pi \varepsilon} + A_3 \frac{b^2}{D}  N^2+\mathcal O(1),   \label{MFLT3d}
\end{align}
\end{subequations}
where $\varepsilon$ is the radius centered at one end, $D$ is the diffusion coefficient, $\kappa=d k_BT/b^2$ is the spring constant with $d$ the spatial dimension, $k_B$ is
the Boltzmann coefficient and $T$ the temperature. The coefficients $A_2$ and $A_3$ (see Fig. \ref{fig:fitting_MFLT} for explicit values) are derived from the second order expansion of the first eigenvalue and weakly depend on $\varepsilon$. We compute them numerically. Using the approximation for end-to-end dynamics of a polymer \cite{Pastor},  the search process was analyzed as a two step process \cite{Toan2008} allowing the authors to postulate Eq. \ref{MFLT3d}, whereas here we derive these scaling laws from considering the polymer configuration space. Although these formulas are derived for fixed $N$ and small $\varepsilon$, we shall see that they are in fact valid for a large range of $N$. Finally, we show that the distribution of the FET can be well approximated by a sum of several exponentials and in most cases a single one is enough. This last result is surprising and has several consequences in understanding the complex chromosomal behavior such as telomere clustering or chromosomal looping inside the nucleus \cite{Ruault2011,Hoze2012}.\\
{\noindent \textit{End-to-end encounter in the configuration space.}-} The stochastic description is that of the Rouse polymer \cite{Doi:Book}, made of a monomer chain at the points $\vect R_n$ ($n=1,2,...,N$) and driven by independent Brownian motions in a force generated by the harmonic potential
\beq\label{eq:Hamiltonian}
\phi(\vect R_{1},..\vect R_{N})_{\rm{Rouse}} = \frac{\kappa}{2} \sum_{n=1}^N \left( \vect R_n - \vect R_{n-1}\right)^2,
\eeq
where $\kappa$ is the spring constant. Each monomer interacts only with its two
neighbors (except for the end points). We neglect all the other
possible interactions such as hydrodynamics and allow the polymer to
cross itself.
\noindent In the Smoluchowski's limit of the Langevin equation
\cite{Schuss:Book}, the dynamics of monomer $\vect R_{n}$ is described by
\beq\label{eq:Langevin_monomer_position}
\frac{d \vect{ R_n}}{d t} = -D\nabla_{\vect{ R_n}} \phi_{\rm{Rouse}} + \sqrt{2D}\frac{d\vect{w_n}}{dt},
\eeq
for $n=1,...,N$, where $\vect w_n$ are independent $d$-dimensional Brownian motions with mean zero and variance $1$.\\
The two ends $\vect R_N,\vect R_1$ meet within a distance of
$\varepsilon<b$:
\beq\label{eq:loop_real}
\left| \vect R_N - \vect R_1 \right| \leq \varepsilon.
\eeq
In the Rouse coordinates, $ \vect{u_p} = \sum^{N}_{n=1} \alpha^n_p \vect R_n$
\cite{Doi:Book}
where
\beq\label{Rousecoef}
\alpha^n_{p}=\left\{
\begin{array}{cc}
\sqrt{\frac{1}{N}},\ &\textrm{} p=0 \\
\sqrt{\frac{2}{N}} \cos \left((n-1/2) \frac{p\pi}{N}\right), &\textrm{ otherwise}
\end{array}%
\right.
\eeq
condition \eqref{eq:loop_real} becomes
\beq\label{S_boundary_condition}
\left| 2\sqrt{\frac{2}{N}} \ds \sum_{p \; \rm{odd}} \vect u_p \cos(p\pi/2N)
\right|\leq \varepsilon.
\eeq
The end-to-end encounter is independent of the coordinate $\vect u_0$, which is the center of mass. Thus, the MFET becomes the mean first passage time for the $(N-1)d$-dimensional stochastic dynamical system
\beq
\vect u(t) =\textbf{(}\vect u_1(t),...,\vect u_{N-1}(t)\textbf{)}
\in \Omega \times \Omega...\times \Omega=\tilde \Omega,
\eeq
where $\Omega=\rR^2$ or $\rR^3$ and
\beq\label{eq:eta_dynamics}
\frac{d \vect u_p}{d t} = -D_p \kappa_p \vect u_p  + \sqrt{2D_p}\frac{ d\vect{  \widetilde{w_p}}}{dt},
\eeq
[$D_p=D,\kappa_p=4\kappa \sin\(p\pi/2N\)^2$ and $p=1,...,N-1$] to the boundary of the
domain $ S_{\epsilon}=\{P \in \tilde \Omega \hbox{ such that } \textrm{dist} (P,{\cal S})\leq
\frac{\varepsilon}{\sqrt{2}} \}$. Each $\vect{\widetilde{w_p}}$ is an independent $d$-dimensional Brownian
motions with mean zero and variance $1$, "dist" is the Euclidean
distance and
\beq\label{defS}
{\cal S}=\{{(\vect u_1,..\vect u_{N-1}) \in \tilde \Omega } \big |\sum_{p \; \rm{odd}} \vect u_p \cos(p\pi/2N) =0\}
\eeq
is a submanifold of codimension $d$ in $\tilde \Omega $. The probability density function (pdf) $p(\vect u(t)=\vect x,t)$
characterizes the dynamics of $\vect u(t)$ and satisfies the forward-Fokker-Planck equation
\cite{Schuss:Book}:
\beq\label{FFP2}
\frac{1}{D}\frac{\partial p(\vect x,t)}{\partial t}&=& \Delta p(\vect x,t) + \nabla\cdot\(  \nabla\phi \:p(\vect x,t)\) = \mathcal{L} p,\nonumber\\
p(\vect x,0)&=&p_0(\vect x), \label{abs_cond}
\eeq
with boundary condition $ p(\vect x,t)=0 $ for $x\in \partial S_{\epsilon}$, $p_0(\vect x)$ is the initial distribution {and $\phi = \frac{1}{2}\sum_{p} \kappa_p \vect u^2_p$} (we shall work in units of $k_BT)$.  The solution of
Eqs. \eqref{FFP2} can be expanded as
\beq\label{eq:eigen_F_FP}
p(\vect x,t) = \ds \sum^{\infty}_{i=0} a_i
w_{\lambda^{\epsilon}_i}( \vect x) e^{-\lambda^{\epsilon}_i tD}
e^{-\phi\left(\vect x
\right)},
\eeq
where $w_{\lambda^{\epsilon}_i}( \vect x)$ and $\lambda^{\epsilon}_i
$ are respectively the eigenfunctions and eigenvalues of the
operator $\mathcal L$ in $\Omega_{\epsilon} = \tilde
\Omega-S_{\epsilon}$ and $a_i$ are coefficients. The probability distribution that the two ends have not met before time $t$ is $p(t) = \textrm{Pr}\{ \tau_{\epsilon}>t\}=\int_{\Omega_{\epsilon} }p(\vect
x,t)dx$, where the first time it happens is
\beq
\tau_\epsilon =\inf\{t>0, \vect u(t) \in \p S_{\epsilon}\}.
\eeq
Using expansion (\ref{eq:eigen_F_FP}), $ p(t)= \sum^{\infty}_{i=0} C_i
e^{-\lambda^{\epsilon}_i D t}$ { where }$ C_i=
\int_{\Omega_{\epsilon}} p_0(\vect x) w_{\lambda^{\epsilon}_i}
(\vect x)d \vect x
\int_{\Omega_{\epsilon}} w_{\lambda^{\epsilon}_i} (\vect x) e^{-\phi\left(\vect x \right)}d \vect
x.$ Starting with an equilibrium distribution $p_0(\vect x) =
|\tilde \Omega|^{-1} e^{-\phi\left(\vect x \right)}$, we have $C_i
=|\tilde \Omega|^{-1} (\int_{\Omega_{\epsilon}}
w_{\lambda^{\epsilon}_i} (\vect x) e^{-\phi\left(\vect x
\right)}d\vect x)^2$ and finally,
\beq\label{MFLT_eigen}
\langle \tau_\epsilon \rangle =\sum^{\infty}_{i=0} \frac{C_i}{D\lambda^{\epsilon}_i}.
\eeq
Our goal is now to estimate the eigenvalues and the coefficients $C_i$. First, for $N$ not too large, a single exponential is sufficient to approximate the FET  [Fig. \ref{fig:merge3d}b] $N=16$ and $32$,
[$p_{N}(t)=\lambda_{N}e^{-\lambda_{N}t}$] and $\varepsilon=0.1b$. Here $\lambda_{16}=
0.0125b^{-2}, \lambda_{32}=0.0063b^{-2}$, while for long polymers, a sum of two exponentials is more accurate to account for
the beginning of the histogram $p_{N}(t)=C_{0}e^{-\lambda^{\epsilon}_0
t}+C_{1}e^{-\lambda^{\epsilon}_1 t}$. For $N=64$, we have
$\lambda^{\epsilon}_0=0.0012b^{-2}, \lambda^{\epsilon}_1=0.0375b^{-2}, C_0=0.99, C_1=0.28$. Although
the two exponential approximation works well for small
$\varepsilon< 0.2b$, we needed four exponents for larger
$\varepsilon$ ($0.4b$). Indeed, for this value, { the series
approximation is less precise.} We use a best fitting procedure to
extract the parameters $\lambda^{\epsilon}_i$ and $C_i$.
Interestingly, for a significant range of $N \in [4-64]$,
$C_0\approx 1$, while $C_1$ remains {approximately} constant for a
given value of $\varepsilon$. For example, for $\varepsilon=0.1b$,
$C_1$ varied with $N$ from $0.2$ to $0.28$ [Fig.
\ref{fig:fitting_C}c]. Interestingly, we observe that
{for} $\varepsilon\uparrow, C_0(\varepsilon)$ is decreasing while
{for} $\varepsilon\uparrow, C_1(\varepsilon)$ is increasing. At this
stage, we conclude that the first two exponentials are sufficient to study the FET and we shall now compute the first eigenvalues $\lambda^{\epsilon}_0$ and $\lambda^{\epsilon}_1$.\\
%%%%%%%%%%%%%%%%%%%%%%%%%%%%%%%%%%%%%%%%%%%%%
{\noindent\it{Estimation of the two first eigenvalues.}}- The
eigenvalues $\lambda^{\epsilon}_i$, $i=0$ and 1, of the operator
$\mathcal{L}$ [Eq.\ref{FFP2}] are obtained by solving the
forward-Fokker-Planck in $\rR^{d(N-1)}$, where the absorbing
boundary is the tubular neighborhood of the $d$-dimensional
submanifold ${\cal S}$. Indeed, for small $\varepsilon$, the perturbation expansion of the eigenvalues is obtained for the Laplace operator with an absorbing boundary condition on the
tubular neighborhood $S_{\epsilon}$ \cite{Chavel1988}, which gives for $d=3,2$, respectively
\beq\label{chavel}
\lambda^{\epsilon}_i&=&\lambda^{0}_{i} +c_2 \epsilon \int_{S} w_{\lambda^{0}_{i}}^2 dV_x+\mathcal O(\epsilon^2), \label{chavel3d}\\
\lambda^{\epsilon}_i&=&\lambda^{0}_{i} + \frac{2\pi}{ \log \epsilon} \int_{S} w_{\lambda^{0}_i}^2 dV_x+\mathcal O\(\(\frac{1}{\log \epsilon}\)^2\), \label{chavel2d}
\eeq
where the eigenfunction $w_{\lambda^{0}_i}$ and eigenvalues $\lambda^0_{i}$ are associated with the nonperturbed operator (no boundary). The volume element $dV_x = e^{-\phi(\vect x)} d\vect x_g$, $d\vect
x_g$ is a measure over the submanifold with $c_2=\frac{2\pi^{3/2}}{\Gamma(3/2)}$ \cite{Chavel1988}. Here the unperturbed eigenfunctions $w_{\lambda^{0}_i}$ are products of Hermite polynomials \cite{Abramowitz1964}, depending on the spatial coordinate, and the eigenvalues $\lambda^{0}_i$ are the sum of one
dimensional eigenvalues obtained in the product. The first eigenfunction associated with the zero eigenvalue is $w_{\lambda^{0}_{0}} = |\tilde \Omega|^{-1/2}$, while the ones associated with the first two modes ($p=1,2$) are ($w_{\lambda^{0}_{1,j}}=\sqrt{\kappa_1}|\tilde \Omega|^{-1/2}
u^j_{1}$) and ($w_{\lambda^{0}_{2,j}}=\sqrt{\kappa_2}|\tilde \Omega|^{-1/2}
u^j_{2}$) with $\lambda^{0}_1 = \kappa_1$ and $\lambda^{0}_2 = \kappa_2$, respectively. \\
The main result here is an explicit computation of the first eigenvalue for small $\varepsilon$. Starting from relation (\ref{chavel3d}) in dimension three with $\lambda^{0}_{i} =0$, we get
\beq\label{firsteigen3d}
\lambda^{\epsilon}_0 = \frac{c_2 \epsilon \ds \int_{S} e^{-\phi\left(\vect x \right) }
 d\vect x_g }{|\tilde \Omega|}+ \mathcal{O(}\epsilon^2 \mathcal{)}.
\eeq
This is the ratio of the closed polymer ensemble to the polymer configuration space.
A straightforward computation with the potential $\phi$ gives $|\tilde \Omega| =
\int_{\Omega} e^{-\phi\left(\vect x \right) }d\vect x_g =
\left[\frac{(2\pi)^{(N-1)}}{\prod_1^{N-1} \kappa_p} \right]^{d/2},$
while using a parametrization of the constraint (\ref{defS}), we get
$ \int_{S} e^{-\phi \left(\vect x \right) }d\vect x_g  =
\left[\frac{(2\pi)^{N-2} \prod_{p \;
\rm{odd}}\omega^2_p}{\prod_{p}\kappa_p\( \sum_{p \; \rm{odd}}
\frac{\omega^2_p}{\kappa_p} \)} \right]^{d/2} $, where $\omega_p =
\cos(p\pi/2N)$. Finally, summarizing these results, using a
direct computation, we obtain that for fixed $N$ and small $\varepsilon$
%\beq\label{first_eigen_eps}
%  \lambda^{\epsilon}_{0} = \left\{
%  \begin{array}{l l}
%     \left( \frac{\kappa}{N\pi}\right)^{3/2}4\pi \epsilon + \mathcal{O}(\epsilon^2) & \quad \text{for } d=3, \\\\
%    \frac{2\kappa}{N\log\left(\frac{\sqrt{2}b}{\varepsilon}\right)} + \mathcal{O}\(\(\frac1{\log\epsilon}\)^2\) & \quad \text{for } d=2.\\
%  \end{array} \right.
%\eeq
\beq\label{first_eigen_eps}
  \lambda^{\epsilon}_{0} = \left\{
  \begin{array}{l l}
     \left( \frac{\kappa}{N\pi}\right)^{3/2}4\pi \epsilon + \mathcal{O}(\epsilon^2) & \quad \text{for } d=3, \\\\
    \frac{2\kappa}{N\log\left(\frac{b}{\epsilon}\right)} + \mathcal{O}\(\(\frac1{\log\epsilon}\)^2\) & \quad \text{for } d=2.\\
  \end{array} \right.
\eeq

This result shows that for small $\varepsilon$, the MFET
depends linearly on $\frac1{\varepsilon}$, confirmed by Brownian
simulations [Fig. \ref{fig:merge3d}(c)]. Using a similar analysis, we
obtain for large $N$ that
\beq
\lambda^{\epsilon}_1    &\approx& \kappa_1 + \epsilon c_2 \(\frac{\kappa}{N\pi}\)^{3/2} \( 1 - \frac{8}{\pi^2}\) +\mathcal{O}(\epsilon^2 ), \label{first_eigen3d_1term}\\
\lambda^{\epsilon}_{2} &\approx& \kappa_2 + \left( \frac{\kappa}{N\pi}\right)^{3/2}4\pi \epsilon + \mathcal{O(}\epsilon^2 ).
\label{second_eigen_eps}
\eeq
Surprisingly, we have
found that the next smallest eigenvalue contributing to the FET is
$\lambda^\epsilon_2$ (affiliated with $w_{\lambda^{0}_{2,j}}$) [Fig. \ref{fig:fitting_C}(a)] but not
$\lambda^\epsilon_1$. All of these estimates are obtained for fixed
$N$ and small $\eps$. When $N$ is not too large, the MFET is well
approximated by $\frac1{\lambda^{\epsilon}_{0}}$. However, when
$\varepsilon$ or $N$ are increasing, more terms are needed in the
expansion of $\lambda^{\epsilon}_{0}$, while the other eigenvalues
do not contribute much, as can be observed in the spectral gap in
the log scale [Fig. \ref{fig:fitting_C}(a)]. This result confirms that the
FET is almost Poissonian. In addition, a direct computation shows
that the second term in the expansion of $\lambda^{\epsilon}_{0}$
is proportional to $1/N$. Because $C_0 \sim1$, using the first
eigenvalue in relation (\ref{MFLT_eigen}), we obtain the approximation
(in dimension 3)
\beq\label{MFET_eign2}
\langle \tau_\varepsilon \rangle_{3d}&\approx&\frac{1}{ D\lambda^{\epsilon}_{0} } =
\frac{1}{D\( \left( \frac{\kappa}{N\pi}\right)^{3/2}4\pi \epsilon -A\epsilon^2/Nb^4\)},
\eeq
where $A$ is a constant and using that $\epsilon=\frac{\varepsilon}{\sqrt{2}}$, we obtain relation (\ref{MFLT3d}). We confirm the validity of the MFET formula
\ref{MFLT2d3d} for a large range of $N$ with Brownian simulations (Fig.\ref{fig:fitting_MFLT}). The value of the coefficient $A_3$, obtained from the fitting procedure is close to the coefficient of $N^2$ term in \cite{Toan2008}, Eq. 13 ($0.053b^2/D$ for our parameters), estimated from the Wilemski-Fixman-approximation method \cite{Wilemski2} of the MFET. \\
\noindent To conclude, we obtain three unpredictable results: First, the FET is
well approximated by a single exponential, showing that the
associated stochastic process is almost Poissonian. Consequently,
modeling cellular biology processes such as nuclear organization,
chromosomes or telomere motion can be well characterized by a single
parameter, instead of using the full polymer dynamics. This
approximation allows us to study telomere clustering in the Yeast
nucleus \cite{Hoze2012}. Another example includes the dynamics of
DNA repair \cite{Gartenberg2009} or the arrival of a DNA fragment to a small target.
Second,  by increasing the radius $\varepsilon$ or the size $N$, the asymptotic for the MFET is obtained by
the other terms in the expansion of the first eigenvalue, but not by
higher order eigenvalues. Two scales are involved in the MFET, one proportional to $N^2$ and the other to $N^{3/2}$, but both are contained in the first eigenvalue and do not arise from {higher} ones. Finally, it is surprising that the regular perturbation of the Fokker-Planck operator in $\varepsilon$ introduces a novel scale with $N$ {in all eigenvalues}. A complete expansion for the MFET has to be found and it would be interesting to derive an exact value for the constants $A_2$ and $A_3$.

{\noindent This research is supported by an ERC-starting Grant.}
\bibliographystyle{apsrev}

\end{document}